\documentclass[twocolumn,pra,aps,superscriptaddress,longbibliography]{revtex4-2}
\usepackage{amsmath}
\usepackage{amssymb}
\usepackage{amsfonts}
\usepackage[dvips]{graphicx}
\usepackage{subfigure}
\usepackage{dcolumn}
\usepackage{txfonts}
\usepackage{bm}
\usepackage{makeidx}
\usepackage{color}
\usepackage{mathtools}
\usepackage{threeparttable}
\usepackage[colorlinks,linkcolor=blue,anchorcolor=blue,citecolor=blue,urlcolor=blue]{hyperref}
\usepackage{lipsum}
\usepackage{braket}
\usepackage{esint}

\begin{document}

\title{Orbital angular momentum and dynamics of off-axis vortex light }

\author{Rui-Feng Liang}
\affiliation{Center for Quantum Sciences and School of Physics, Northeast Normal University, Changchun 130024, China}

\author{L. Yang}
\affiliation{Center for Quantum Sciences and School of Physics, Northeast Normal University, Changchun 130024, China}

\begin{abstract}
The orbital angular momentum (OAM) of light and optical vortices are closely related concepts that are often conflated. The conserved OAM arises fundamentally from the SO(3) rotational symmetry of spacetime, while the concept of vortices originates from fluid mechanics. In this work, we investigate the OAM of off-axis vortex light to clarify the distinction between the two concepts. We also examine the propagation of vortex beams, revealing the dynamic behavior of both the off-axis vortex center and photon flux within the transverse plane. This helps us explore the fundamental differences between the OAM quantum number and the vortex topological charge (TC).\end{abstract}

\maketitle
\section{Introduction}
In free space, both the spin and orbital angular momentum (OAM) of light are conserved due to the SO(3) rotational symmetry of spacetime~\cite{greiner2013field, cohen1997photons}. As early as 1936, Bethe experimentally discovered and measured the spin angular momentum of light along propagating axis~\cite{beth1936mechnical}. However, it was not until 1992 that Allen et al. demonstrated that Laguerre-Gaussian beams with helical wavefronts carry orbital angular momentum~\cite{Allen1992Orbital,beijersbergen1993astigmatic}. Over the past three decades, the orbital angular momentum of photons has garnered significant attention, particularly due to its broad applications in quantum information~\cite{yao2011orbital}. By encoding information in higher dimensions using OAM states, it is possible to surpass the theoretical limits of traditional two-dimensional polarization encoding methods~\cite{Mair2001Entanglement, barreiro2008beating}. Recently, advancements in precise control technologies for the phase, intensity, and polarization of light fields—such as spatial light modulators and metasurfaces—have expanded the applications of OAM beyond quantum information, branching into interdisciplinary domains. For example, the torque characteristics of OAM have been utilized to facilitate non-contact rotation~\cite{padgett2011tweezers, Dholakia2011shaping, simpson1997mechanical} and three-dimensional trapping of microparticles~\cite{Gahagan1996optical}. Furthermore, leveraging the phase-sensitive properties of OAM light has led to innovations in super-resolution microscopy techniques~\cite{bernet2006quantitative}. Additionally, high-dimensional quantum entangled states have been constructed to enhance the parallelism of qubit operations~\cite{babazadeh2017high}, and highly sensitive molecular conformation analysis has been achieved through the interactions between OAM modes and biomolecules~\cite{zhuang2004unraveling}, among various other applications.

The concept of optical vortices was first introduced by Coullet et al. in 1989, drawing inspiration from fluid mechanics. They derived a paraxial beam solution that featured a helical phase~\cite{Coullet1989optical}. Notably, the filed intensity at the center of an vortex is zero~\cite{shen2019optical, Yihua2022Vortexbeam, Guo2022Generation, molina2007twisted}. The generation of vortex beams has evolved significantly through various developmental stages, progressing from static optical components~\cite{Alison2011Optical} to dynamically tunable systems~\cite{ma2017generation}, and transitioning from macroscopic optical configurations to micro/nano-integrated photonic platforms~\cite{Nanfang2011Light}. Furthermore, high-purity vortex beams are now generated on-chip using silicon-based waveguides and microring resonators~\cite{Xuewen2018Recent}. At the same time, optical vortex lattices—two- or three-dimensional arrays of multiple OAM beams arranged in specific crystallographic symmetries (such as hexagonal or Kagome lattices)—have emerged as a rapidly developing frontier~\cite{Liuhao2021Optical}. These vortex arrays offer unique advantages in parallel optical manipulation, high-dimensional optical communications, and topological photonics~\cite{Du2024Optical}.

Previous investigations of vortex beams have revealed conceptual ambiguities between optical OAM and photon vortices, particularly with regard to the conflation of OAM quantum numbers and topological charge (TC). Non-helical light traveling along curved fibers can carry non-zero OAM, while symmetrically distributed anti-vortex dipole beams can result in vanishing OAM along the propagation axis. Crucially, OAM has an inherent vector nature, where only its axial component along the propagation direction directly correlates with photonic vorticity. Additionally, the concept of a vortex is rooted in hydrodynamic systems and is fundamentally characterized by photon flux density and its associated circulation. This study clarifies the distinctions between OAM and photonic vortices by analyzing all three components of the OAM in off-axis vortex beams. The results show that the mean OAM per photon generally deviates from integer multiples of $\hbar$, whereas the TC of the vortex beam remains quantized in integer values. Moreover, we demonstrate that both the vortex core and the photon flux centroid follow a straight-line trajectory in the transverse plane during beam propagation. Notably, the TC remains a conserved quantity throughout this propagation process.

\section{Research on the Dynamics of Off-Axis Vortex Beams in the Transverse Plane}

We begin by investigating the dynamic evolution of multi-vortex Gaussian beams during propagation. 
Consider a scalar light field characterized by $N$ off-axis vortices, all possessing identical topological charge signs. The field function at the focal plane ($z = 0$) is represented as a spiral function $[x+\mathrm{i}y-(x_{n,0}+\mathrm{i}y_{n,0})]^{m_{n}}$ combined with a Gaussian function that has a beam waist of $w_0$ \cite{Guy1993Optical}. Here, $m_n$ denotes the topological charge associated with the $n$-th vortex, while $(x_{n,0}, y_{n,0})$ indicates its initial position in the focal plane. The beam can be expressed as a linear superposition of on-axis vortex fields. By employing the Angular Spectrum Representation method~\cite{Novotny2012principles}, we compute the propagation characteristics of each axial vortex field component. Ultimately, we derive the propagation field for the off-axis vortex beam \cite{Guy1993Optical}.
\begin{align}
 & \xi\left(x,y,z\right)\nonumber \\
= & \frac{1}{\mathcal{N}}\left[\frac{w_{0}^{2}}{w^{2}\left(z\right)}\right]\exp\left[-\frac{x^{2}+y^{2}}{w^{2}\left(z\right)}\right]\nonumber \\
 & \times\prod_{n=1}^{N}\left[w_{0}\frac{\left(x\pm\mathrm{i}y\right)-\left(x_{n}\pm\mathrm{i}y_{n}\right)(1+\mathrm{i}\frac{z}{z_{R}})}{w^{2}\left(z\right)}\right]^{\left|m_{n}\right|},\label{1}
\end{align}
where $z$ is the propagation distance, $z_{R}=\pi w_{0}^{2}/{\lambda}$ is the Rayleigh distance, $\lambda$ is the wavelength, $w\left(z\right)=w_{0}\sqrt{1+\mathrm{i}z/z_{R}}$, and $1/{\mathcal{N}}$ is the normalization factor.

\begin{figure}[h]
\includegraphics[width=8cm]{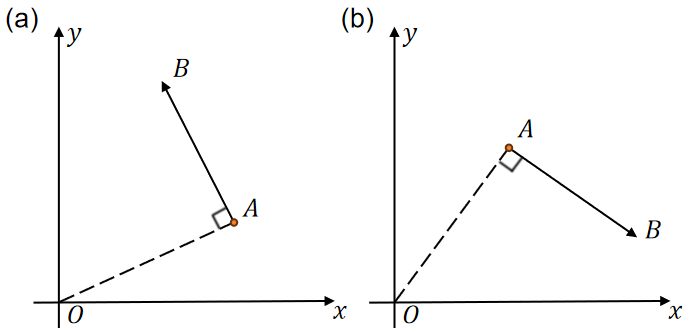}
\caption{\label{Fig1}
The trajectory of the $n$-th off-axis vortex center as it propagates within the transverse plane}
\end{figure}

It can be observed that during the propagation process, the position of each vortex center is determined by the zero point of the complex amplitude function:
\begin{equation}
\left(x\pm \mathrm{i}y\right)-\left(x_{n}\pm \mathrm{i}y_{n}\right)(1+\mathrm{i}z/z_{R})=0. 
\end{equation}
In the focal plane, the position of the 
$n$-th vortex center is given by the complex vector $\vec{\rho}_{n,0}=x_{n,0}+\mathrm{i}y_{n,0}$.
As the vortex beam propagates, the position of the vortex center will shift to $\vec{\rho}_{n}=\vec{\rho}_{n,0}(1+\mathrm{i}z/z_{R})$. When a complex vector is multiplied by the imaginary number$\mathrm{i}$, it will rotate counterclockwise by $90^\circ$ in the complex plane. Therefore, during the propagation of the beam, the vortex center will move along the direction of a straight line perpendicular to the vector $\vec{\rho}_{n}$ in the transverse plane. As shown in Fig. \ref{Fig1}, for a vortex with a positive topological charge, its center will move in the counterclockwise direction, while the center of a vortex with a negative topological charge will move in the clockwise direction. When the beam propagates to infinity, the vortex center will rotate by $90^\circ$\cite{Guy1993Optical}.

\section{The Photon Flux and Topological Charge of Off-axis Vortex Beams}

For an extended period, phase singularities have been a central focus in the study of photon vortices~\cite{Berry2004optical}. The concept of vortices is rooted in fluid mechanics, and it is essential to give due consideration to the associated observable physical quantities. By introducing the photon effective field operator $\hat{\psi}(\vec{r})$ into the classical paraxial Helmholtz equation, one can derive the continuity equation that photons satisfy in a plane perpendicular to their propagation direction~\cite{Yang2022Quantum}. Consequently, we can define the photon number density (PND) operator $\hat{\psi}^{\dagger}(\vec{r})\hat{\psi}(\vec{r})$ as well as the corresponding photon current density operator 
$\hat{j}(\vec{r})=(-\mathrm{i}/2k_{0})\{ \hat{\psi}^{\dagger}(\vec{r})\vec{\nabla}_{T}\hat{\psi}(\vec{r})-[\vec{\nabla}_{T}\hat{\psi}^{\dagger}(\vec{r})]\hat{\psi}(\vec{r})\}$, where $k_0$ represents the wave number corresponding to the central frequency of the quasi-monochromatic beam. Here, $\vec{\nabla}_{T}$ denotes the differential operator in the transverse plane defined as $\vec{e}_{x}\frac{\partial}{\partial x} + \vec{e}_{y}\frac{\partial}{\partial y}$, with $\vec{e}_{i}(i=x,y)$ being unit vectors. Ultimately, by integrating this framework with descriptions of quantum states for photon pulses, one can develop quantum theories pertaining to vortex optical fields and other structured optical fields~\cite{yang2021non}.

\begin{figure}[h]
\includegraphics[width=9cm]{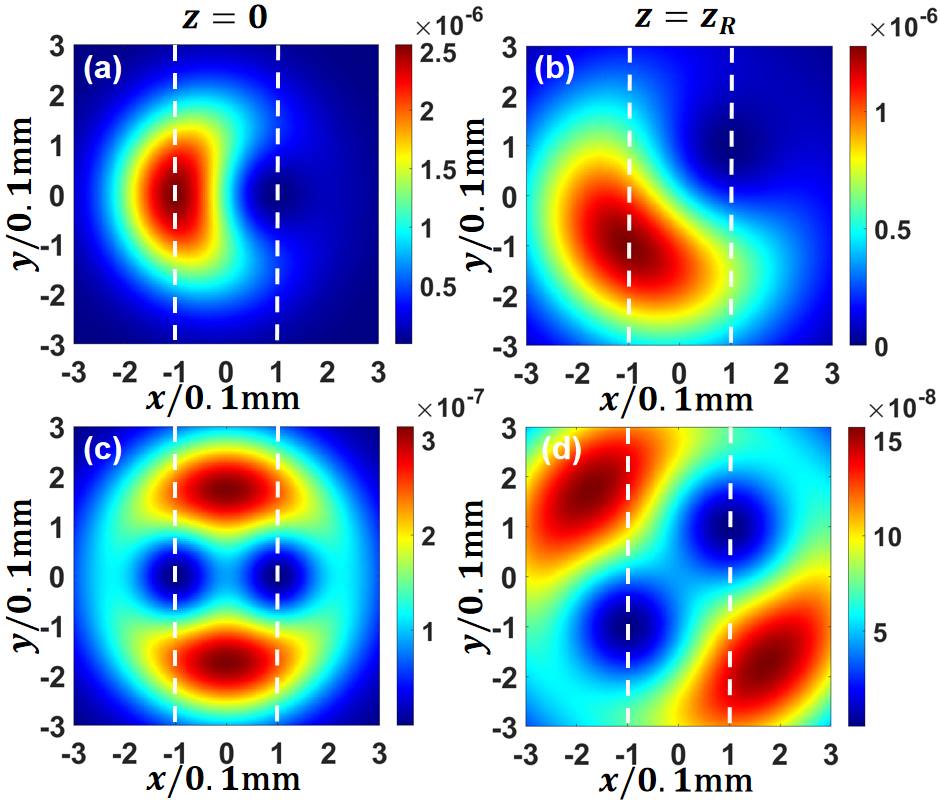}
\caption{\label{Fig2}
Photon number density profile of off-axis vortex beams during propagation}
\end{figure}

\subsection{The Photon Flux of Off-axis Vortex Beams}

A laser beam can be described by a coherent state $|\alpha_{\xi}\rangle$~\cite{Yang2022Quantum}, where the strength of the laser beam is characterized by the amplitude~$\alpha$, and the shape of the beam is described by Eq.~(\ref{1}). By averaging the PND in this coherent state, we derive that $\langle \hat{\psi}^{\dagger}(\vec{r})\hat{\psi}(\vec{r})\rangle/|\alpha|^2=|\xi (\vec{r})|^2$. In this study, a Gaussian beam with a wavelength of $632.8$ nm and a beam waist of $w_{0}=0.2$ mm is considered. is considered. In the experiments, the Gaussian beam can be transformed into an off-axis single vortex beam with a topological charge of $+1$ at $[0.5w_{0},0]$, or an off-axis double vortex beam with centers at $[\pm 0.5w_{0},0]$ using a spatial light modulator (SLM).
In Fig. \ref{Fig2}, the dynamic behavior of PND as the beam propagates is illustrated. The normalization reference unit for dimensional standardization is defined using $ w_{0} $, with a specification that 1 unit in the image coordinate system corresponds to an actual physical displacement of 0.1 mm. Figures \ref{Fig2}(a) and (b) present the distributions of PND for the off-axis single vortex beam at both the focal plane and at $ z=z_{R} $ (where $ z_{R}\approx200 $ mm). It can be observed that the phase of the wave packet function cannot be determined at points where PND reaches zero, and it will be demonstrated later that photon flow velocity diverges at these locations.
As the beam propagates, its image rotates counterclockwise around the origin. At the beam waist, diffraction effects lead to significant lateral diffusion of the beam's cross-section. Essentially, during propagation, while the center of a single vortex moves upward along a straight line defined by perpendicular complex vector $ \vec{\rho}_{1,0}=(1,0) $, regions exhibiting high PND shift downward along the negative direction of the $ y $-axis.
Figures \ref{Fig2}(c) and (d) respectively show the PND distributions of the off-axis double vortex beam at the focal plane and the plane of $z=z_{R}$. Throughout this propagation process, similar counterclockwise rotation occurs across all images. The vortex centers initially located at $\vec{\rho}_{1,0}=(1,0)$ and $\vec{\rho}_{2,0}=(-1,0)$ move along the two white dashed lines($x=\pm1$)to the positions of $\vec{\rho}_{1}=(1,1)$ and $\vec{\rho}_{2}=(-1,-1)$ respectively.
\begin{figure}[h]
\includegraphics[width=7.7cm]{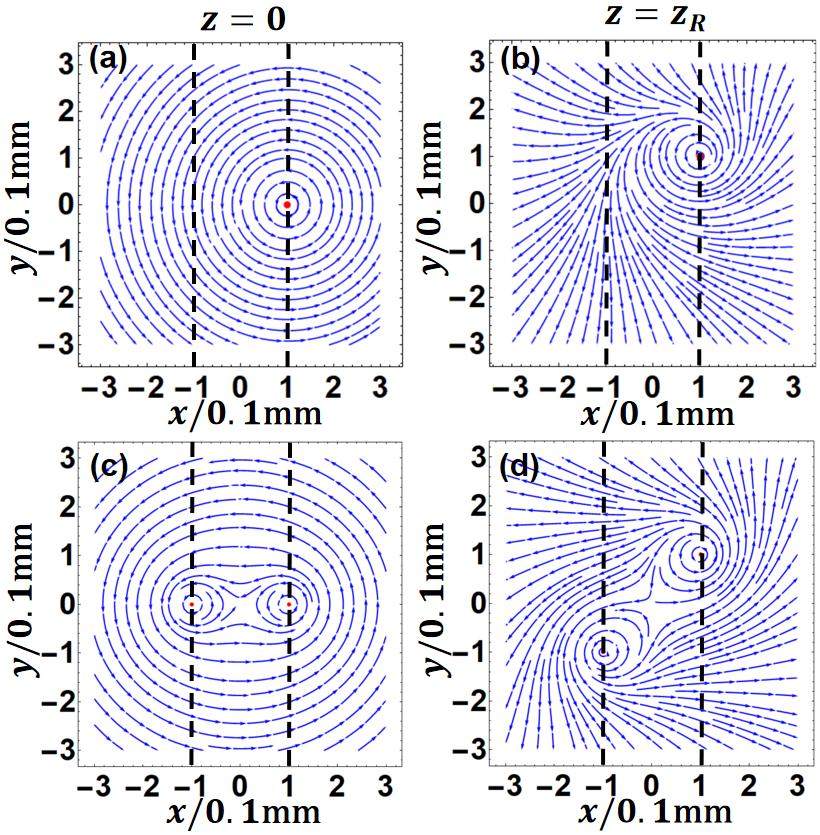}
\caption{\label{Fig3}
PND profile of off-axis vortex beams during propagation}
\end{figure}

Similarly, by calculating the average of the photon current density operator in the coherent state, the components of the expected value of the photon current density of the vortex beam can be obtained~\cite{Yang2022Quantum}
\begin{equation}
\langle \hat{j}_{i}\left(\vec{r}\right)\rangle =-\frac{\mathrm{i}}{2k_{0}}\left\{ \xi^{*}\left(\vec{r}\right)\partial_i\xi\left(\vec{r}\right)-\left[\partial_i\xi^{*}\left(\vec{r}\right)\right]\xi\left(\vec{r}\right)\right\} \vec{e}_{i}.\label{3}
\end{equation}
In Fig. \ref{Fig3}, the dynamic behavior of the photon current as the beam propagates is illustrated. Figures \ref{Fig3}(a) and (b) respectively show the photon current images of the off-axis single vortex beam with a TC of $m=+1$ at the focal plane and the plane of 
$z=z_{R}$. It can be seen that the photon current rotates counterclockwise around the positive vortex center. Figures \ref{Fig3} (c) and (d) respectively display the photon current images of the off-axis double vortex beam with a TC of $m=+1$ at the focal plane and the plane of $z=z_{R}$.  In this case, the photon current encircles two moving positive vortex centers, resulting in two counterclockwise rotating vortices. Conversely, if both topological charges are negative, two clockwise rotating vortices will emerge. It is worth noting that from Fig. \ref{Fig3} (a) to (b) or from Fig. \ref{Fig3} (c) to (d), that is, during the beam propagation process, the spatial distribution of the photon flow no longer exhibits the standard circular symmetry feature. This is due to the significant radial component generated by the photon flow during the wavefront evolution process.

\subsection{The Conservation of the Topological Charge of Off-axis Single Vortex Beams}

Based on the photon number density and photon current density operators, the local photon flow velocity can be defined as $v(\vec{r})=\langle \hat{j}(\vec{r})\rangle /\langle \hat{n}(\vec{r})\rangle$ and the circulation along a closed path as $\kappa=\ointctrclockwise_{T}v\left(\vec{r}\right)\cdot d\vec{r}$.
The quantum number corresponding to the quantized circulation is equal to the TC number. Therefore, the topological charge of the vortex is defined as the winding number of the circulation of the vortex optical field~\cite{Yang2022Quantum}. By constructing the circulation along the closed path around the vortex center, a criterion for the conservation of the topological charge of the off-axis vortex is established: if $\kappa(z)=\kappa(0)$ is satisfied under the condition of any propagation distance $z$, it proves that the topological charge remains strictly conserved during the propagation of the beam. 

In the cylindrical coordinate system, the scalar field function of an off-axis single vortex with a topological charge of $m$ in the focal plane can be written as the product of a helical function $[\rho e^{\mathrm{i}\theta}-\rho_{1}e^{(\mathrm{i}\theta_{1})}]^{m}$and a Gaussian function with a beam waist of $w_0$~\cite{Guy1993Optical}. Through the translation of the coordinate system 
$[\rho e^{\mathrm{i}\theta}-\rho_{1}e^{\mathrm{i}\theta_{1}}]^{m}=[\rho^{\prime}e^{\mathrm{i}\theta^{\prime}}]^{m}$. By constructing a new coordinate system $O^{\prime}(\rho^{\prime},\theta^{\prime})$with the position of the off-axis single vortex center$(\rho_{1},\theta_{1})$as the origin, the off-axis single vortex beam can be transformed into an on-axis vortex beam. Then, the scalar field function of the single vortex beam in the new coordinate system $O^{\prime}\left(\rho^{\prime},\theta^{\prime}\right)$ is reconstructed as
\begin{align}
 & \xi\left(\rho^{\prime},\theta^{\prime},0\right)\nonumber \\
= & \exp\left\{-\frac{\rho_{1}^{2}+\rho^{\prime2}+2\rho_{1}\rho^{\prime}\cos\left(\theta^{\prime}-\theta_{1}\right)}{w_{0}^{2}}\right\}\left(\frac{\rho^{\prime}}{w_{0}}\right)^{m}e^{im\theta^{\prime}}.\label{4}
\end{align}
First, according to Eq. (\ref{3}), the expected value of the photon current density is calculated as $\langle \hat{j}(\vec{r}^{\prime})\rangle =m\xi^{2}(\rho^{\prime},\theta^{\prime})/k_{0}\rho^{\prime}\vec{e}_{\theta^{\prime}}$,
where
$\xi(\rho^{\prime},\theta^{\prime})=(\rho^{\prime}/w_{0})^{m}\exp[-(\rho_{1}^{2}+\rho^{\prime2}+2\rho_{1}\rho^{\prime}\cos(\theta^{\prime}-\theta_{1}))/w_{0}^{2}]$
is real. It can be seen that in the focal plane, the photon current has only a tangential component, which is the reason why the spatial distribution of the photon current in Fig. \ref{Fig3} (a) exhibits a circular symmetry characteristic. Second, the velocity of the corresponding photon current is:
$v(\vec{r}^{\prime})=m/k_{0}\rho^{\prime}\vec{e}_{\theta^{\prime}}$, By performing a circular closed-loop integral of the velocity around the vortex center (ensuring that 
$d\rho^{\prime}=0$),  the circulation is calculated to be $\kappa(0)=2\pi\cdot m/k_{0}$.

 Similarly, according to Eq. (\ref{1}), through the translation of the coordinate system $[\rho e^{\mathrm{i}\theta}-\rho_{1}e^{\mathrm{i}\theta_{1}}(1+\mathrm{i}z/z_{R})]^{m}=[\rho^{\prime}e^{\mathrm{i}\theta^{\prime}}]^{m}$, the propagation field of the single vortex beam in the new coordinate system$O^{\prime}(\rho^{\prime},\theta^{\prime})$can be expressed as:
\begin{align}
 & \xi\left(\rho^{\prime},\theta^{\prime},z\right)\nonumber \\
= & \frac{1}{\mathcal{N}}\left[\frac{\rho^{\prime}e^{\mathrm{i}\theta^{\prime}}}{w_{0}}\right]^{m}\left(\frac{1}{1+\mathrm{i}\frac{z}{z_{R}}}\right)^{m+1}\exp\left\{ -\frac{\rho_{1}^{2}(1+\frac{z^{2}}{z_{R}^{2}})+\rho^{\prime2}}{w_{0}^{2}(1+\mathrm{i}\frac{z}{z_{R}})}\right\} \nonumber \\
 & \times\exp\left\{ -\frac{2\rho_{1}\rho^{\prime}[\cos(\theta_{1}-\theta^{\prime})-\sin(\theta_{1}-\theta^{\prime})\frac{z}{z_{R}}]}{w_{0}^{2}(1+\mathrm{i}\frac{z}{z_{R}})}\right\}.\label{13}
\end{align}
By implementing the above solution steps (for the focal plane), the circulation of the single vortex beam at the plane with an arbitrary propagation distance $z$ is calculated, and it is found that $\kappa(z)=\kappa(0)$. This finding demonstrates that the topological charge of the off-axis single vortex beam is conserved throughout the propagation process.

\section{The Orbital Angular Momentum of Off-axis Vortex Beams}

In recent years, researchers have developed a quantum theory of the spin and orbital angular momentum of optical fields based on quantum field theory and Noether's theorem~\cite{yang2022Quantumfield}. They have also rewritten the observable orbital angular momentum operator of photons using the photon effective field operator ~\cite{yang2021non}
\begin{equation}
\hat{L}=\int d^{3}r\hat{\psi}^{\dagger}\left(\vec{r}\right)\left(\vec{r}\times\hat{p}\right)\hat{\psi}\left(\vec{r}\right)\equiv\int d^{3}r\hat{\psi}^{\dagger}\left(\vec{r}\right)\hat{\mathfrak{l}}\hat{\psi}\left(\vec{r}\right).\label{6}
\end{equation}
The quantum state of an off-axis vortex beam can be expressed 
$\left|\psi\right\rangle =\sum_{n}\mathcal{C}_{n}\left|\Psi_{n}\left(\vec{r}\right)\right\rangle$. By taking the average of the photon orbital angular momentum operator (\ref{6}) in this quantum state, the components of the expected value of the orbital angular momentum can be calculated
\begin{align}
\langle \hat{L}_{i}\rangle = &  \sum_{nn^{\prime}}\mathcal{C}_{n}^{*}\mathcal{C}_{n^{\prime}}\langle \Psi_{n}(\vec{r})|\hat{L}_{i}|\Psi_{n^{\prime}}\left(\vec{r}\right)\rangle,\label{7} 
\end{align}
the variables 
$i=x,y,z$ represent the components of the coordinate system. $\Psi_{n}\left(\vec{r}\right)$ as the component of the on-axis vortex field, is normalized, and $\mathcal{C}_{n}$ denotes the corresponding expansion coefficient. 

\begin{figure}[h]
\includegraphics[width=8cm]{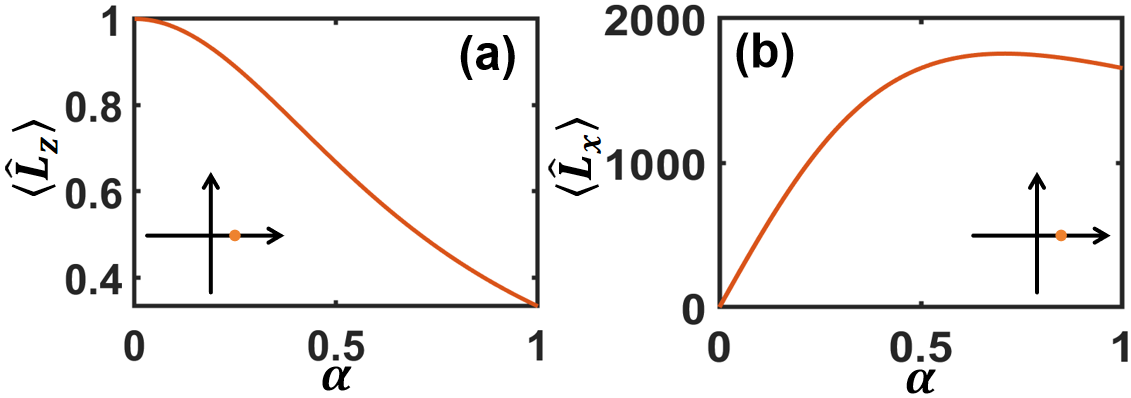}
\caption{\label{Fig4}
Numerical distribution plots of $\langle \hat{L}_{z}\rangle$ and $\langle \hat{L}_{x}\rangle$ in unit of $\hbar$ for an off-axis single vortex beam as $\alpha$ varies.}
\end{figure}

Taking a single vortex beam with a topological charge of $+1$ and a center at $w_{0}=$0.2~mm as an example, we derive the mean OAM $\langle\hat{L}_{z}\rangle =\hbar w_{0}^{2}/[2(x_{1}^{2}+y_{1}^{2})+w_{0}^{2}]$ according to Eq.~(\ref{7}), where $(x_{1},y_{1})$ is the transverse coordinates of the vortex center. It can be seen that when the vortex is located on the optical axis, $\langle \hat{L}_{z}\rangle=\hbar$, and the quantum number of the orbital angular momentum strictly corresponds to the topological charge number, which is in line with the characteristics of the angular momentum eigenstate in quantum mechanics. When the vortex center deviates from the optical axis, the off-axis vortex field can always be written as a superposition state of the on-axis vortex field components, making $\langle\hat{L}_{z}\rangle$ no longer an integer multiple of $\hbar$. In Fig. \ref{Fig4} (a), the off-axis single vortex center is placed at the position 
$[\alpha w_{0},0]$ ($\alpha\in[0,1]$) on the 
$x$-axis, and a numerical distribution diagram of $\langle\hat{L}_{z}\rangle$ varying with $\alpha$ is plotted. The results indicate that the expectation value $\langle\hat{L}_{z}\rangle$ of the off-axis single vortex beam decreases with the increase of  $\alpha$. The magnitude is associated with the geometric position of the vortex, rather than being determined by the intrinsic property of the topological charge number.

To evaluate the mean values of $\hat{L}_{x}$ and $\hat{L}_{y}$ for a continuous wave beam, a spatial cut-off along the z-axis must be applied. This results in that $\hat{L}_{x}$ and $\hat{L}_{y}$ no longer being Hermitian operators, which means their expected values in the coherent state are no longer real numbers. To address this issue, the mean values
$\langle \hat{L}_{x}\rangle$ and $\langle \hat{L}_{y}\rangle$ are evaluated in this paper using the following relations
\begin{align}
\langle \hat{L}_{x}\rangle = & \frac{1}{2}\{\langle \hat{L}_{x}\rangle _{n<n^{\prime}}+\langle \hat{L}_{x}\rangle _{n>n^{\prime}}\},\label{8} \\
\langle \hat{L}_{y}\rangle = & \frac{1}{2}\{ \langle \hat{L}_{y}\rangle _{n<n^{\prime}}+\langle \hat{L}_{y}\rangle _{n>n^{\prime}}\}.\label{9} 
\end{align} 
According to Eqs. (\ref{8}) and (\ref{9}), the obtained $\langle\hat{L}_{x}\rangle$ and $\langle \hat{L}_{y}\rangle$ of the off-axis single vortex beam are
\begin{align}
\langle \hat{L}_{x}\rangle = &  \frac{5(w_{0}^{2} + 2z_{R}^{2})x_{1} + w_{0}^{2}y_{1}}{z_{R}[4(x_{1}^{2} + y_{1}^{2}) + 2w_{0}^{2}]}\hbar,\label{10} \\
\langle \hat{L}_{y}\rangle = & \frac{5(w_{0}^{2} + 2z_{R}^{2})y_{1} - w_{0}^{2}x_{1}}{z_{R}[4(x_{1}^{2} + y_{1}^{2}) + 2w_{0}^{2}]}\hbar.\label{11} 
\end{align} 
When the vortex is located on the optical axis, the rotational symmetry of the system leads to $\langle \hat{L}_{x}\rangle=\langle \hat{L}_{y}\rangle=0$. When the vortex deviates from the optical axis, the values of 
$\langle \hat{L}_{x}\rangle$ and $\langle \hat{L}_{y}\rangle$ are determined by the position of the vortex, and they are generally no longer integer multiples of $\hbar$. In Fig. \ref{Fig4}(b), the off-axis vortex center is placed at the position $[\alpha w_{0},0]$ on the $x$-axis, and a numerical distribution diagram of 
$\langle\hat{L}_{x}\rangle$ of the off-axis single vortex varying with $\alpha $ is plotted. The data curve exhibits a non-monotonic evolution characteristic: $\langle \hat{L}_{x}\rangle$ first reaches a maximum value and then gradually decays as $\alpha$ increases. This behavior arises because, once the vortex configuration is established, the ring-shaped PND in the transverse plane is also determined accordingly, reflecting both the light intensity distribution and the probability of photon occurrence. Within the optimal annular region, the probability of high light intensity is relatively elevated; conversely, when moving away from this region, there is a significant decrease in light intensity, with zero intensity observed at the vortex center. Consequently, $\langle \hat{L}_{x}\rangle$ first increases and then decreases as $\alpha$ rises. If the vortex is placed at the position $[0,\alpha w_{0}]$ on the $y$-axis, the numerical distribution of $\langle \hat{L}_{y}\rangle$is exactly the same as that in Fig. \ref{Fig4}(b), indicating that $\langle\hat{L}_{x}\rangle$ and $\langle\hat{L}_{y}\rangle$ possess rotational symmetry. However, the values of $\langle\hat{L}_{x}\rangle$ and $\langle\hat{L}_{y}\rangle$ are relatively large, which is mainly due to the small wavelength.

\begin{figure}[h]
\includegraphics[width=8cm]{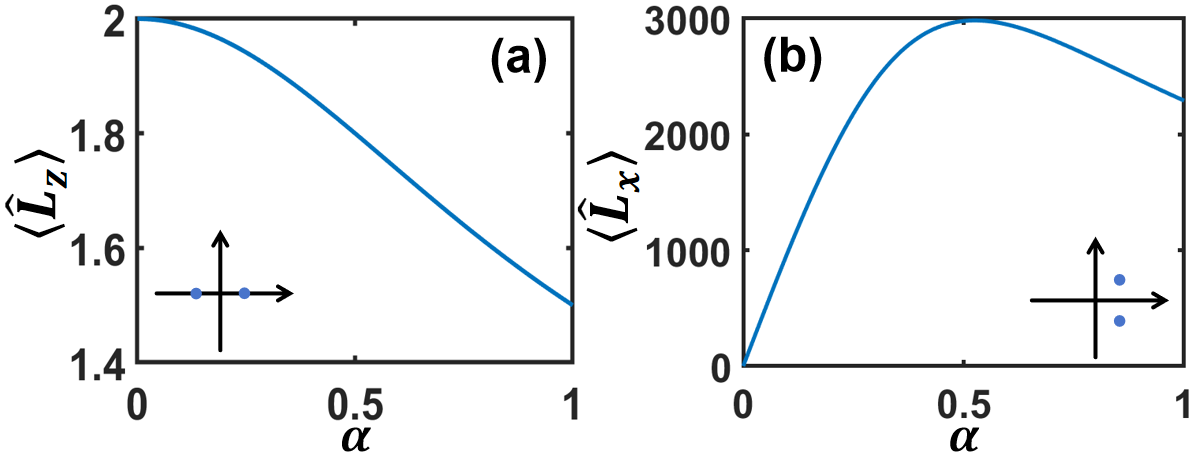}
\caption{\label{Fig5}
Numerical distribution plots of $\langle \hat{L}_{z}\rangle$ and $\langle \hat{L}_{x}\rangle$ for off-axis double vortex beams with same topological charge signs as $\alpha$ varies.}
\end{figure}

Similarly, we consider an off-axis double vortex beam with beam waist $w_{0}=$0.2 mm and topological charges $m_{1}=m_{2}=+1$. Firstly, when the off-axis double vortices are positioned at $[\pm\alpha w_{0},0]$on the 
$x$-axis, according to Eq. (\ref{7}), $\langle\hat{L}_{z}\rangle =2\hbar/(1+2\alpha^{4})$. In Fig.~\ref{Fig5} (a), a numerical distribution graph of $\langle \hat{L}_{z}\rangle$ varying with $\alpha$under this condition is plotted. It can be observed that when the double vortices are on the axis ($\alpha=0$), $\langle \hat{L}_{z}\rangle=2\hbar$. However, when the two vortices are off the axis, 
$\langle\hat{L}_{z}\rangle$ is no longer an integer multiple of $\hbar$ and gradually decreases as $\alpha$ increases. Secondly, when the off-axis double vortices are symmetrically placed about the 
$x$-axis at positions$[\alpha w_{0},\pm\alpha w_{0}]$ through calculation, we obtain
\begin{equation}
\langle \hat{L}_{x}\rangle =\frac{5(3\alpha+4\alpha^{3})w_{0}^{2}+20(\alpha+2\alpha^{3})z_{R}^{2}}{2z_{R}\{ 1+4\alpha^{2}+8\alpha^{4}\} w_{0}}.\label{12}
\end{equation}
In Fig.~\ref{Fig5}(b), a numerical distribution graph of $\langle \hat{L}_{x}\rangle$ as $\alpha$ changes under this condition is plotted, which also exhibits a distribution pattern of increasing first and then decreasing.

\section{CONCLUSION}
Based on the propagation field theory of off-axis vortex beams in real space, we investigates their dynamic characteristics in the transverse plane. By analyzing the propagation dependence of the PND distribution, we found that the vortex center moves along a straight line perpendicular to the line connecting the origin and the vortex center. We calculated the mean value of the photon current density operator in the transverse plane, the local flow velocity, and the circulation to characterize the spatial distribution of the photon current. This approach allowed us to rigorously prove the conservation of the topological charge of off-axis single vortex beams during propagation. Next, by calculating and analyzing the expected values of the orbital angular momentum components of off-axis vortex beams, we found that the average orbital angular momentum carried by each photon is generally not an integer multiple of $\hbar$. However, the topological charges of these beams still strictly maintain the quantized property of integer invariance. Our examples demonstrate that the key parameter of the quantized topological charge can be defined and characterized independently of the quantum number of orbital angular momentum, highlighting the difference between the OAM and photon vortices.

\bibliography{main}

\begin{thebibliography}{32}%
\makeatletter
\providecommand \@ifxundefined [1]{%
 \@ifx{#1\undefined}
}%
\providecommand \@ifnum [1]{%
 \ifnum #1\expandafter \@firstoftwo
 \else \expandafter \@secondoftwo
 \fi
}%
\providecommand \@ifx [1]{%
 \ifx #1\expandafter \@firstoftwo
 \else \expandafter \@secondoftwo
 \fi
}%
\providecommand \natexlab [1]{#1}%
\providecommand \enquote  [1]{``#1''}%
\providecommand \bibnamefont  [1]{#1}%
\providecommand \bibfnamefont [1]{#1}%
\providecommand \citenamefont [1]{#1}%
\providecommand \href@noop [0]{\@secondoftwo}%
\providecommand \href [0]{\begingroup \@sanitize@url \@href}%
\providecommand \@href[1]{\@@startlink{#1}\@@href}%
\providecommand \@@href[1]{\endgroup#1\@@endlink}%
\providecommand \@sanitize@url [0]{\catcode `\\12\catcode `\$12\catcode
  `\&12\catcode `\#12\catcode `\^12\catcode `\_12\catcode `\%12\relax}%
\providecommand \@@startlink[1]{}%
\providecommand \@@endlink[0]{}%
\providecommand \url  [0]{\begingroup\@sanitize@url \@url }%
\providecommand \@url [1]{\endgroup\@href {#1}{\urlprefix }}%
\providecommand \urlprefix  [0]{URL }%
\providecommand \Eprint [0]{\href }%
\providecommand \doibase [0]{https://doi.org/}%
\providecommand \selectlanguage [0]{\@gobble}%
\providecommand \bibinfo  [0]{\@secondoftwo}%
\providecommand \bibfield  [0]{\@secondoftwo}%
\providecommand \translation [1]{[#1]}%
\providecommand \BibitemOpen [0]{}%
\providecommand \bibitemStop [0]{}%
\providecommand \bibitemNoStop [0]{.\EOS\space}%
\providecommand \EOS [0]{\spacefactor3000\relax}%
\providecommand \BibitemShut  [1]{\csname bibitem#1\endcsname}%
\let\auto@bib@innerbib\@empty
\bibitem [{\citenamefont {Greiner}\ and\ \citenamefont
  {Reinhardt}(2013)}]{greiner2013field}%
  \BibitemOpen
  \bibfield  {author} {\bibinfo {author} {\bibfnamefont {W.}~\bibnamefont
  {Greiner}}\ and\ \bibinfo {author} {\bibfnamefont {J.}~\bibnamefont
  {Reinhardt}},\ }\href@noop {} {\emph {\bibinfo {title} {Field
  quantization}}}\ (\bibinfo  {publisher} {Springer Science \& Business
  Media},\ \bibinfo {year} {2013})\BibitemShut {NoStop}%
\bibitem [{\citenamefont {Cohen-Tannoudji}\ \emph {et~al.}(1997)\citenamefont
  {Cohen-Tannoudji}, \citenamefont {Dupont-Roc},\ and\ \citenamefont
  {Grynberg}}]{cohen1997photons}%
  \BibitemOpen
  \bibfield  {author} {\bibinfo {author} {\bibfnamefont {C.}~\bibnamefont
  {Cohen-Tannoudji}}, \bibinfo {author} {\bibfnamefont {J.}~\bibnamefont
  {Dupont-Roc}},\ and\ \bibinfo {author} {\bibfnamefont {G.}~\bibnamefont
  {Grynberg}},\ }\href@noop {} {\emph {\bibinfo {title} {Photons and
  Atoms-Introduction to Quantum Electrodynamics}}}\ (\bibinfo  {publisher}
  {Wiley-VCH},\ \bibinfo {year} {1997})\BibitemShut {NoStop}%
\bibitem [{\citenamefont {Beth}(1936)}]{beth1936mechnical}%
  \BibitemOpen
  \bibfield  {author} {\bibinfo {author} {\bibfnamefont {R.~A.}\ \bibnamefont
  {Beth}},\ }\bibfield  {title} {\bibinfo {title} {Mechanical detection and
  measurement of the angular momentum of light},\ }\href
  {https://doi.org/10.1103/PhysRev.50.115} {\bibfield  {journal} {\bibinfo
  {journal} {Phys. Rev.}\ }\textbf {\bibinfo {volume} {50}},\ \bibinfo {pages}
  {115} (\bibinfo {year} {1936})}\BibitemShut {NoStop}%
\bibitem [{\citenamefont {Allen}\ \emph {et~al.}(1992)\citenamefont {Allen},
  \citenamefont {Beijersbergen}, \citenamefont {Spreeuw},\ and\ \citenamefont
  {Woerdman}}]{Allen1992Orbital}%
  \BibitemOpen
  \bibfield  {author} {\bibinfo {author} {\bibfnamefont {L.}~\bibnamefont
  {Allen}}, \bibinfo {author} {\bibfnamefont {M.~W.}\ \bibnamefont
  {Beijersbergen}}, \bibinfo {author} {\bibfnamefont {R.~J.~C.}\ \bibnamefont
  {Spreeuw}},\ and\ \bibinfo {author} {\bibfnamefont {J.~P.}\ \bibnamefont
  {Woerdman}},\ }\bibfield  {title} {\bibinfo {title} {Orbital angular momentum
  of light and the transformation of laguerre-gaussian laser modes},\ }\href
  {https://doi.org/10.1103/PhysRevA.45.8185} {\bibfield  {journal} {\bibinfo
  {journal} {Phys. Rev. A}\ }\textbf {\bibinfo {volume} {45}},\ \bibinfo
  {pages} {8185} (\bibinfo {year} {1992})}\BibitemShut {NoStop}%
\bibitem [{\citenamefont {Beijersbergen}\ \emph {et~al.}(1993)\citenamefont
  {Beijersbergen}, \citenamefont {Allen}, \citenamefont {Van~der Veen},\ and\
  \citenamefont {Woerdman}}]{beijersbergen1993astigmatic}%
  \BibitemOpen
  \bibfield  {author} {\bibinfo {author} {\bibfnamefont {M.~W.}\ \bibnamefont
  {Beijersbergen}}, \bibinfo {author} {\bibfnamefont {L.}~\bibnamefont
  {Allen}}, \bibinfo {author} {\bibfnamefont {H.}~\bibnamefont {Van~der
  Veen}},\ and\ \bibinfo {author} {\bibfnamefont {J.}~\bibnamefont
  {Woerdman}},\ }\bibfield  {title} {\bibinfo {title} {Astigmatic laser mode
  converters and transfer of orbital angular momentum},\ }\href@noop {}
  {\bibfield  {journal} {\bibinfo  {journal} {Optics Communications}\ }\textbf
  {\bibinfo {volume} {96}},\ \bibinfo {pages} {123} (\bibinfo {year}
  {1993})}\BibitemShut {NoStop}%
\bibitem [{\citenamefont {Yao}\ and\ \citenamefont
  {Padgett}(2011{\natexlab{a}})}]{yao2011orbital}%
  \BibitemOpen
  \bibfield  {author} {\bibinfo {author} {\bibfnamefont {A.~M.}\ \bibnamefont
  {Yao}}\ and\ \bibinfo {author} {\bibfnamefont {M.~J.}\ \bibnamefont
  {Padgett}},\ }\bibfield  {title} {\bibinfo {title} {Orbital angular momentum:
  origins, behavior and applications},\ }\href
  {https://doi.org/10.1364/AOP.3.000161} {\bibfield  {journal} {\bibinfo
  {journal} {Advances in optics and photonics}\ }\textbf {\bibinfo {volume}
  {3}},\ \bibinfo {pages} {161} (\bibinfo {year}
  {2011}{\natexlab{a}})}\BibitemShut {NoStop}%
\bibitem [{\citenamefont {Mair}\ \emph {et~al.}(2001)\citenamefont {Mair},
  \citenamefont {Vaziri}, \citenamefont {Weihs},\ and\ \citenamefont
  {Zeilinger}}]{Mair2001Entanglement}%
  \BibitemOpen
  \bibfield  {author} {\bibinfo {author} {\bibfnamefont {A.}~\bibnamefont
  {Mair}}, \bibinfo {author} {\bibfnamefont {A.}~\bibnamefont {Vaziri}},
  \bibinfo {author} {\bibfnamefont {G.}~\bibnamefont {Weihs}},\ and\ \bibinfo
  {author} {\bibfnamefont {A.}~\bibnamefont {Zeilinger}},\ }\bibfield  {title}
  {\bibinfo {title} {Entanglement of the orbital angular momentum states of
  photons},\ }\href@noop {} {\bibfield  {journal} {\bibinfo  {journal}
  {Nature}\ }\textbf {\bibinfo {volume} {412}},\ \bibinfo {pages} {313}
  (\bibinfo {year} {2001})}\BibitemShut {NoStop}%
\bibitem [{\citenamefont {Barreiro}\ \emph {et~al.}(2008)\citenamefont
  {Barreiro}, \citenamefont {Wei},\ and\ \citenamefont
  {Kwiat}}]{barreiro2008beating}%
  \BibitemOpen
  \bibfield  {author} {\bibinfo {author} {\bibfnamefont {J.~T.}\ \bibnamefont
  {Barreiro}}, \bibinfo {author} {\bibfnamefont {T.-C.}\ \bibnamefont {Wei}},\
  and\ \bibinfo {author} {\bibfnamefont {P.~G.}\ \bibnamefont {Kwiat}},\
  }\bibfield  {title} {\bibinfo {title} {Beating the channel capacity limit for
  linear photonic superdense coding},\ }\href@noop {} {\bibfield  {journal}
  {\bibinfo  {journal} {Nature physics}\ }\textbf {\bibinfo {volume} {4}},\
  \bibinfo {pages} {282} (\bibinfo {year} {2008})}\BibitemShut {NoStop}%
\bibitem [{\citenamefont {Padgett}\ and\ \citenamefont
  {Bowman}(2011)}]{padgett2011tweezers}%
  \BibitemOpen
  \bibfield  {author} {\bibinfo {author} {\bibfnamefont {M.}~\bibnamefont
  {Padgett}}\ and\ \bibinfo {author} {\bibfnamefont {R.}~\bibnamefont
  {Bowman}},\ }\bibfield  {title} {\bibinfo {title} {Tweezers with a twist},\
  }\href@noop {} {\bibfield  {journal} {\bibinfo  {journal} {Nature photonics}\
  }\textbf {\bibinfo {volume} {5}},\ \bibinfo {pages} {343} (\bibinfo {year}
  {2011})}\BibitemShut {NoStop}%
\bibitem [{\citenamefont {Dholakia}\ and\ \citenamefont
  {{\v{C}}i{\v{z}}m{\'a}r}(2011)}]{Dholakia2011shaping}%
  \BibitemOpen
  \bibfield  {author} {\bibinfo {author} {\bibfnamefont {K.}~\bibnamefont
  {Dholakia}}\ and\ \bibinfo {author} {\bibfnamefont {T.}~\bibnamefont
  {{\v{C}}i{\v{z}}m{\'a}r}},\ }\bibfield  {title} {\bibinfo {title} {Shaping
  the future of manipulation},\ }\href@noop {} {\bibfield  {journal} {\bibinfo
  {journal} {Nature photonics}\ }\textbf {\bibinfo {volume} {5}},\ \bibinfo
  {pages} {335} (\bibinfo {year} {2011})}\BibitemShut {NoStop}%
\bibitem [{\citenamefont {Simpson}\ \emph {et~al.}(1997)\citenamefont
  {Simpson}, \citenamefont {Dholakia}, \citenamefont {Allen},\ and\
  \citenamefont {Padgett}}]{simpson1997mechanical}%
  \BibitemOpen
  \bibfield  {author} {\bibinfo {author} {\bibfnamefont {N.}~\bibnamefont
  {Simpson}}, \bibinfo {author} {\bibfnamefont {K.}~\bibnamefont {Dholakia}},
  \bibinfo {author} {\bibfnamefont {L.}~\bibnamefont {Allen}},\ and\ \bibinfo
  {author} {\bibfnamefont {M.}~\bibnamefont {Padgett}},\ }\bibfield  {title}
  {\bibinfo {title} {Mechanical equivalence of spin and orbital angular
  momentum of light: an optical spanner},\ }\href@noop {} {\bibfield  {journal}
  {\bibinfo  {journal} {Optics letters}\ }\textbf {\bibinfo {volume} {22}},\
  \bibinfo {pages} {52} (\bibinfo {year} {1997})}\BibitemShut {NoStop}%
\bibitem [{\citenamefont {Gahagan}\ and\ \citenamefont
  {Swartzlander~Jr}(1996)}]{Gahagan1996optical}%
  \BibitemOpen
  \bibfield  {author} {\bibinfo {author} {\bibfnamefont {K.}~\bibnamefont
  {Gahagan}}\ and\ \bibinfo {author} {\bibfnamefont {G.}~\bibnamefont
  {Swartzlander~Jr}},\ }\bibfield  {title} {\bibinfo {title} {Optical vortex
  trapping of particles},\ }\href@noop {} {\bibfield  {journal} {\bibinfo
  {journal} {Optics Letters}\ }\textbf {\bibinfo {volume} {21}},\ \bibinfo
  {pages} {827} (\bibinfo {year} {1996})}\BibitemShut {NoStop}%
\bibitem [{\citenamefont {Bernet}\ \emph {et~al.}(2006)\citenamefont {Bernet},
  \citenamefont {Jesacher}, \citenamefont {F{\"u}rhapter}, \citenamefont
  {Maurer},\ and\ \citenamefont {Ritsch-Marte}}]{bernet2006quantitative}%
  \BibitemOpen
  \bibfield  {author} {\bibinfo {author} {\bibfnamefont {S.}~\bibnamefont
  {Bernet}}, \bibinfo {author} {\bibfnamefont {A.}~\bibnamefont {Jesacher}},
  \bibinfo {author} {\bibfnamefont {S.}~\bibnamefont {F{\"u}rhapter}}, \bibinfo
  {author} {\bibfnamefont {C.}~\bibnamefont {Maurer}},\ and\ \bibinfo {author}
  {\bibfnamefont {M.}~\bibnamefont {Ritsch-Marte}},\ }\bibfield  {title}
  {\bibinfo {title} {Quantitative imaging of complex samples by spiral phase
  contrast microscopy},\ }\href@noop {} {\bibfield  {journal} {\bibinfo
  {journal} {Optics Express}\ }\textbf {\bibinfo {volume} {14}},\ \bibinfo
  {pages} {3792} (\bibinfo {year} {2006})}\BibitemShut {NoStop}%
\bibitem [{\citenamefont {Babazadeh}\ \emph {et~al.}(2017)\citenamefont
  {Babazadeh}, \citenamefont {Erhard}, \citenamefont {Wang}, \citenamefont
  {Malik}, \citenamefont {Nouroozi}, \citenamefont {Krenn},\ and\ \citenamefont
  {Zeilinger}}]{babazadeh2017high}%
  \BibitemOpen
  \bibfield  {author} {\bibinfo {author} {\bibfnamefont {A.}~\bibnamefont
  {Babazadeh}}, \bibinfo {author} {\bibfnamefont {M.}~\bibnamefont {Erhard}},
  \bibinfo {author} {\bibfnamefont {F.}~\bibnamefont {Wang}}, \bibinfo {author}
  {\bibfnamefont {M.}~\bibnamefont {Malik}}, \bibinfo {author} {\bibfnamefont
  {R.}~\bibnamefont {Nouroozi}}, \bibinfo {author} {\bibfnamefont
  {M.}~\bibnamefont {Krenn}},\ and\ \bibinfo {author} {\bibfnamefont
  {A.}~\bibnamefont {Zeilinger}},\ }\bibfield  {title} {\bibinfo {title}
  {High-dimensional single-photon quantum gates: concepts and experiments},\
  }\href@noop {} {\bibfield  {journal} {\bibinfo  {journal} {Physical review
  letters}\ }\textbf {\bibinfo {volume} {119}},\ \bibinfo {pages} {180510}
  (\bibinfo {year} {2017})}\BibitemShut {NoStop}%
\bibitem [{\citenamefont {Zhuang}(2004)}]{zhuang2004unraveling}%
  \BibitemOpen
  \bibfield  {author} {\bibinfo {author} {\bibfnamefont {X.}~\bibnamefont
  {Zhuang}},\ }\bibfield  {title} {\bibinfo {title} {Unraveling dna
  condensation with optical tweezers},\ }\href@noop {} {\bibfield  {journal}
  {\bibinfo  {journal} {Science}\ }\textbf {\bibinfo {volume} {305}},\ \bibinfo
  {pages} {188} (\bibinfo {year} {2004})}\BibitemShut {NoStop}%
\bibitem [{\citenamefont {Coullet}\ \emph {et~al.}(1989)\citenamefont
  {Coullet}, \citenamefont {Gil},\ and\ \citenamefont
  {Rocca}}]{Coullet1989optical}%
  \BibitemOpen
  \bibfield  {author} {\bibinfo {author} {\bibfnamefont {P.}~\bibnamefont
  {Coullet}}, \bibinfo {author} {\bibfnamefont {L.}~\bibnamefont {Gil}},\ and\
  \bibinfo {author} {\bibfnamefont {F.}~\bibnamefont {Rocca}},\ }\bibfield
  {title} {\bibinfo {title} {Optical vortices},\ }\href@noop {} {\bibfield
  {journal} {\bibinfo  {journal} {Optics Communications}\ }\textbf {\bibinfo
  {volume} {73}},\ \bibinfo {pages} {403} (\bibinfo {year} {1989})}\BibitemShut
  {NoStop}%
\bibitem [{\citenamefont {Shen}\ \emph {et~al.}(2019)\citenamefont {Shen},
  \citenamefont {Wang}, \citenamefont {Xie}, \citenamefont {Min}, \citenamefont
  {Fu}, \citenamefont {Liu}, \citenamefont {Gong},\ and\ \citenamefont
  {Yuan}}]{shen2019optical}%
  \BibitemOpen
  \bibfield  {author} {\bibinfo {author} {\bibfnamefont {Y.}~\bibnamefont
  {Shen}}, \bibinfo {author} {\bibfnamefont {X.}~\bibnamefont {Wang}}, \bibinfo
  {author} {\bibfnamefont {Z.}~\bibnamefont {Xie}}, \bibinfo {author}
  {\bibfnamefont {C.}~\bibnamefont {Min}}, \bibinfo {author} {\bibfnamefont
  {X.}~\bibnamefont {Fu}}, \bibinfo {author} {\bibfnamefont {Q.}~\bibnamefont
  {Liu}}, \bibinfo {author} {\bibfnamefont {M.}~\bibnamefont {Gong}},\ and\
  \bibinfo {author} {\bibfnamefont {X.}~\bibnamefont {Yuan}},\ }\bibfield
  {title} {\bibinfo {title} {Optical vortices 30 years on: Oam manipulation
  from topological charge to multiple singularities},\ }\href@noop {}
  {\bibfield  {journal} {\bibinfo  {journal} {Light: Science \& Applications}\
  }\textbf {\bibinfo {volume} {8}},\ \bibinfo {pages} {90} (\bibinfo {year}
  {2019})}\BibitemShut {NoStop}%
\bibitem [{\citenamefont {Bai}\ \emph {et~al.}(2022)\citenamefont {Bai},
  \citenamefont {Lv}, \citenamefont {Fu},\ and\ \citenamefont
  {Yang}}]{Yihua2022Vortexbeam}%
  \BibitemOpen
  \bibfield  {author} {\bibinfo {author} {\bibfnamefont {Y.}~\bibnamefont
  {Bai}}, \bibinfo {author} {\bibfnamefont {H.}~\bibnamefont {Lv}}, \bibinfo
  {author} {\bibfnamefont {X.}~\bibnamefont {Fu}},\ and\ \bibinfo {author}
  {\bibfnamefont {Y.}~\bibnamefont {Yang}},\ }\bibfield  {title} {\bibinfo
  {title} {Vortex beam: generation and detection of orbital angular momentum
  [invited]},\ }\href {https://www.researching.cn/articles/OJ5f21de78f2a1ad65}
  {\bibfield  {journal} {\bibinfo  {journal} {Chinese Optics Letters}\ }\textbf
  {\bibinfo {volume} {20}},\ \bibinfo {pages} {012601} (\bibinfo {year}
  {2022})}\BibitemShut {NoStop}%
\bibitem [{\citenamefont {Guo}\ \emph {et~al.}(2022)\citenamefont {Guo},
  \citenamefont {Chang}, \citenamefont {Meng}, \citenamefont {An},
  \citenamefont {Jia}, \citenamefont {Zhao}, \citenamefont {Wang},\ and\
  \citenamefont {Zhang}}]{Guo2022Generation}%
  \BibitemOpen
  \bibfield  {author} {\bibinfo {author} {\bibfnamefont {Z.}~\bibnamefont
  {Guo}}, \bibinfo {author} {\bibfnamefont {Z.}~\bibnamefont {Chang}}, \bibinfo
  {author} {\bibfnamefont {J.}~\bibnamefont {Meng}}, \bibinfo {author}
  {\bibfnamefont {M.}~\bibnamefont {An}}, \bibinfo {author} {\bibfnamefont
  {J.}~\bibnamefont {Jia}}, \bibinfo {author} {\bibfnamefont {Z.}~\bibnamefont
  {Zhao}}, \bibinfo {author} {\bibfnamefont {X.}~\bibnamefont {Wang}},\ and\
  \bibinfo {author} {\bibfnamefont {P.}~\bibnamefont {Zhang}},\ }\bibfield
  {title} {\bibinfo {title} {Generation of perfect optical vortex by
  laguerre--gauss beams with a high-order radial index},\ }\href
  {https://doi.org/10.1364/AO.461251} {\bibfield  {journal} {\bibinfo
  {journal} {Appl. Opt.}\ }\textbf {\bibinfo {volume} {61}},\ \bibinfo {pages}
  {5269} (\bibinfo {year} {2022})}\BibitemShut {NoStop}%
\bibitem [{\citenamefont {Molina-Terriza}\ \emph {et~al.}(2007)\citenamefont
  {Molina-Terriza}, \citenamefont {Torres},\ and\ \citenamefont
  {Torner}}]{molina2007twisted}%
  \BibitemOpen
  \bibfield  {author} {\bibinfo {author} {\bibfnamefont {G.}~\bibnamefont
  {Molina-Terriza}}, \bibinfo {author} {\bibfnamefont {J.~P.}\ \bibnamefont
  {Torres}},\ and\ \bibinfo {author} {\bibfnamefont {L.}~\bibnamefont
  {Torner}},\ }\bibfield  {title} {\bibinfo {title} {Twisted photons},\
  }\href@noop {} {\bibfield  {journal} {\bibinfo  {journal} {Nature physics}\
  }\textbf {\bibinfo {volume} {3}},\ \bibinfo {pages} {305} (\bibinfo {year}
  {2007})}\BibitemShut {NoStop}%
\bibitem [{\citenamefont {Yao}\ and\ \citenamefont
  {Padgett}(2011{\natexlab{b}})}]{Alison2011Optical}%
  \BibitemOpen
  \bibfield  {author} {\bibinfo {author} {\bibfnamefont {A.~M.}\ \bibnamefont
  {Yao}}\ and\ \bibinfo {author} {\bibfnamefont {M.~J.}\ \bibnamefont
  {Padgett}},\ }\bibfield  {title} {\bibinfo {title} {Orbital angular momentum:
  origins, behavior and applications},\ }\href
  {https://doi.org/10.1364/AOP.3.000161} {\bibfield  {journal} {\bibinfo
  {journal} {Adv. Opt. Photon.}\ }\textbf {\bibinfo {volume} {3}},\ \bibinfo
  {pages} {161} (\bibinfo {year} {2011}{\natexlab{b}})}\BibitemShut {NoStop}%
\bibitem [{\citenamefont {Ma}\ \emph {et~al.}(2017)\citenamefont {Ma},
  \citenamefont {Li}, \citenamefont {Tai}, \citenamefont {Li}, \citenamefont
  {Wang}, \citenamefont {Tang}, \citenamefont {Tang}, \citenamefont {Wang},\
  and\ \citenamefont {Nie}}]{ma2017generation}%
  \BibitemOpen
  \bibfield  {author} {\bibinfo {author} {\bibfnamefont {H.}~\bibnamefont
  {Ma}}, \bibinfo {author} {\bibfnamefont {X.}~\bibnamefont {Li}}, \bibinfo
  {author} {\bibfnamefont {Y.}~\bibnamefont {Tai}}, \bibinfo {author}
  {\bibfnamefont {H.}~\bibnamefont {Li}}, \bibinfo {author} {\bibfnamefont
  {J.}~\bibnamefont {Wang}}, \bibinfo {author} {\bibfnamefont {M.}~\bibnamefont
  {Tang}}, \bibinfo {author} {\bibfnamefont {J.}~\bibnamefont {Tang}}, \bibinfo
  {author} {\bibfnamefont {Y.}~\bibnamefont {Wang}},\ and\ \bibinfo {author}
  {\bibfnamefont {Z.}~\bibnamefont {Nie}},\ }\bibfield  {title} {\bibinfo
  {title} {Generation of circular optical vortex array},\ }\href@noop {}
  {\bibfield  {journal} {\bibinfo  {journal} {Annalen der Physik}\ }\textbf
  {\bibinfo {volume} {529}},\ \bibinfo {pages} {1700285} (\bibinfo {year}
  {2017})}\BibitemShut {NoStop}%
\bibitem [{\citenamefont {Yu}\ \emph {et~al.}(2011)\citenamefont {Yu},
  \citenamefont {Genevet}, \citenamefont {Kats}, \citenamefont {Aieta},
  \citenamefont {Tetienne}, \citenamefont {Capasso},\ and\ \citenamefont
  {Gaburro}}]{Nanfang2011Light}%
  \BibitemOpen
  \bibfield  {author} {\bibinfo {author} {\bibfnamefont {N.}~\bibnamefont
  {Yu}}, \bibinfo {author} {\bibfnamefont {P.}~\bibnamefont {Genevet}},
  \bibinfo {author} {\bibfnamefont {M.~A.}\ \bibnamefont {Kats}}, \bibinfo
  {author} {\bibfnamefont {F.}~\bibnamefont {Aieta}}, \bibinfo {author}
  {\bibfnamefont {J.-P.}\ \bibnamefont {Tetienne}}, \bibinfo {author}
  {\bibfnamefont {F.}~\bibnamefont {Capasso}},\ and\ \bibinfo {author}
  {\bibfnamefont {Z.}~\bibnamefont {Gaburro}},\ }\bibfield  {title} {\bibinfo
  {title} {Light propagation with phase discontinuities: Generalized laws of
  reflection and refraction},\ }\href {https://doi.org/10.1126/science.1210713}
  {\bibfield  {journal} {\bibinfo  {journal} {Science}\ }\textbf {\bibinfo
  {volume} {334}},\ \bibinfo {pages} {333} (\bibinfo {year} {2011})},\ \Eprint
  {https://arxiv.org/abs/https://www.science.org/doi/pdf/10.1126/science.1210713}
  {https://www.science.org/doi/pdf/10.1126/science.1210713} \BibitemShut
  {NoStop}%
\bibitem [{\citenamefont {Wang}\ \emph {et~al.}(2018)\citenamefont {Wang},
  \citenamefont {Nie}, \citenamefont {Liang}, \citenamefont {Wang},
  \citenamefont {Li},\ and\ \citenamefont {Jia}}]{Xuewen2018Recent}%
  \BibitemOpen
  \bibfield  {author} {\bibinfo {author} {\bibfnamefont {X.}~\bibnamefont
  {Wang}}, \bibinfo {author} {\bibfnamefont {Z.}~\bibnamefont {Nie}}, \bibinfo
  {author} {\bibfnamefont {Y.}~\bibnamefont {Liang}}, \bibinfo {author}
  {\bibfnamefont {J.}~\bibnamefont {Wang}}, \bibinfo {author} {\bibfnamefont
  {T.}~\bibnamefont {Li}},\ and\ \bibinfo {author} {\bibfnamefont
  {B.}~\bibnamefont {Jia}},\ }\bibfield  {title} {\bibinfo {title} {Recent
  advances on optical vortex generation},\ }\href
  {https://doi.org/doi:10.1515/nanoph-2018-0072} {\bibfield  {journal}
  {\bibinfo  {journal} {Nanophotonics}\ }\textbf {\bibinfo {volume} {7}},\
  \bibinfo {pages} {1533} (\bibinfo {year} {2018})}\BibitemShut {NoStop}%
\bibitem [{\citenamefont {Zhu}\ \emph {et~al.}(2021)\citenamefont {Zhu},
  \citenamefont {Tang}, \citenamefont {Li}, \citenamefont {Tai},\ and\
  \citenamefont {Li}}]{Liuhao2021Optical}%
  \BibitemOpen
  \bibfield  {author} {\bibinfo {author} {\bibfnamefont {L.}~\bibnamefont
  {Zhu}}, \bibinfo {author} {\bibfnamefont {M.}~\bibnamefont {Tang}}, \bibinfo
  {author} {\bibfnamefont {H.}~\bibnamefont {Li}}, \bibinfo {author}
  {\bibfnamefont {Y.}~\bibnamefont {Tai}},\ and\ \bibinfo {author}
  {\bibfnamefont {X.}~\bibnamefont {Li}},\ }\bibfield  {title} {\bibinfo
  {title} {Optical vortex lattice: an exploitation of orbital angular
  momentum},\ }\href {https://doi.org/doi:10.1515/nanoph-2021-0139} {\bibfield
  {journal} {\bibinfo  {journal} {Nanophotonics}\ }\textbf {\bibinfo {volume}
  {10}},\ \bibinfo {pages} {2487} (\bibinfo {year} {2021})}\BibitemShut
  {NoStop}%
\bibitem [{\citenamefont {Du}\ \emph {et~al.}(2024)\citenamefont {Du},
  \citenamefont {Quan}, \citenamefont {Li},\ and\ \citenamefont
  {Wang}}]{Du2024Optical}%
  \BibitemOpen
  \bibfield  {author} {\bibinfo {author} {\bibfnamefont {J.}~\bibnamefont
  {Du}}, \bibinfo {author} {\bibfnamefont {Z.}~\bibnamefont {Quan}}, \bibinfo
  {author} {\bibfnamefont {K.}~\bibnamefont {Li}},\ and\ \bibinfo {author}
  {\bibfnamefont {J.}~\bibnamefont {Wang}},\ }\bibfield  {title} {\bibinfo
  {title} {Optical vortex array: generation and applications},\ }\href@noop {}
  {\bibfield  {journal} {\bibinfo  {journal} {Chinese Optics Letters}\ }\textbf
  {\bibinfo {volume} {22}},\ \bibinfo {pages} {020011} (\bibinfo {year}
  {2024})}\BibitemShut {NoStop}%
\bibitem [{\citenamefont {Indebetouw}(1993)}]{Guy1993Optical}%
  \BibitemOpen
  \bibfield  {author} {\bibinfo {author} {\bibfnamefont {G.}~\bibnamefont
  {Indebetouw}},\ }\bibfield  {title} {\bibinfo {title} {Optical vortices and
  their propagation},\ }\href {https://doi.org/10.1080/09500349314550101}
  {\bibfield  {journal} {\bibinfo  {journal} {Journal of Modern Optics}\
  }\textbf {\bibinfo {volume} {40}},\ \bibinfo {pages} {73} (\bibinfo {year}
  {1993})},\ \Eprint
  {https://arxiv.org/abs/https://doi.org/10.1080/09500349314550101}
  {https://doi.org/10.1080/09500349314550101} \BibitemShut {NoStop}%
\bibitem [{\citenamefont {Novotny}\ and\ \citenamefont
  {Hecht}(2012)}]{Novotny2012principles}%
  \BibitemOpen
  \bibfield  {author} {\bibinfo {author} {\bibfnamefont {L.}~\bibnamefont
  {Novotny}}\ and\ \bibinfo {author} {\bibfnamefont {B.}~\bibnamefont
  {Hecht}},\ }\href@noop {} {\emph {\bibinfo {title} {Principles of
  nano-optics}}}\ (\bibinfo  {publisher} {Cambridge university press},\
  \bibinfo {year} {2012})\BibitemShut {NoStop}%
\bibitem [{\citenamefont {Berry}(2004)}]{Berry2004optical}%
  \BibitemOpen
  \bibfield  {author} {\bibinfo {author} {\bibfnamefont {M.~V.}\ \bibnamefont
  {Berry}},\ }\bibfield  {title} {\bibinfo {title} {Optical vortices evolving
  from helicoidal integer and fractional phase steps},\ }\href
  {https://doi.org/10.1088/1464-4258/6/2/018} {\bibfield  {journal} {\bibinfo
  {journal} {Journal of Optics A: Pure and Applied Optics}\ }\textbf {\bibinfo
  {volume} {6}},\ \bibinfo {pages} {259} (\bibinfo {year} {2004})}\BibitemShut
  {NoStop}%
\bibitem [{\citenamefont {Yang}\ and\ \citenamefont
  {Xu}(2022)}]{Yang2022Quantum}%
  \BibitemOpen
  \bibfield  {author} {\bibinfo {author} {\bibfnamefont {L.-P.}\ \bibnamefont
  {Yang}}\ and\ \bibinfo {author} {\bibfnamefont {D.}~\bibnamefont {Xu}},\
  }\bibfield  {title} {\bibinfo {title} {Quantum theory of photonic vortices
  and quantum statistics of twisted photons},\ }\href
  {https://doi.org/10.1103/PhysRevA.105.023723} {\bibfield  {journal} {\bibinfo
   {journal} {Phys. Rev. A}\ }\textbf {\bibinfo {volume} {105}},\ \bibinfo
  {pages} {023723} (\bibinfo {year} {2022})}\BibitemShut {NoStop}%
\bibitem [{\citenamefont {Yang}\ and\ \citenamefont
  {Jacob}(2021)}]{yang2021non}%
  \BibitemOpen
  \bibfield  {author} {\bibinfo {author} {\bibfnamefont {L.-P.}\ \bibnamefont
  {Yang}}\ and\ \bibinfo {author} {\bibfnamefont {Z.}~\bibnamefont {Jacob}},\
  }\bibfield  {title} {\bibinfo {title} {Non-classical photonic spin texture of
  quantum structured light},\ }\href@noop {} {\bibfield  {journal} {\bibinfo
  {journal} {Communications Physics}\ }\textbf {\bibinfo {volume} {4}},\
  \bibinfo {pages} {221} (\bibinfo {year} {2021})}\BibitemShut {NoStop}%
\bibitem [{\citenamefont {Yang}\ \emph {et~al.}(2022)\citenamefont {Yang},
  \citenamefont {Khosravi},\ and\ \citenamefont
  {Jacob}}]{yang2022Quantumfield}%
  \BibitemOpen
  \bibfield  {author} {\bibinfo {author} {\bibfnamefont {L.-P.}\ \bibnamefont
  {Yang}}, \bibinfo {author} {\bibfnamefont {F.}~\bibnamefont {Khosravi}},\
  and\ \bibinfo {author} {\bibfnamefont {Z.}~\bibnamefont {Jacob}},\ }\bibfield
   {title} {\bibinfo {title} {Quantum field theory for spin operator of the
  photon},\ }\href@noop {} {\bibfield  {journal} {\bibinfo  {journal} {Physical
  Review Research}\ }\textbf {\bibinfo {volume} {4}},\ \bibinfo {pages}
  {023165} (\bibinfo {year} {2022})}\BibitemShut {NoStop}%
\end{thebibliography}%
\end{document}